\documentclass[12pt]{iopart}

\usepackage{graphicx}
\begin{document}

\title{Saltatory drift in a randomly driven two-wave potential}

\author{G.Oshanin$^{1,2}$, J.Klafter$^3$ and M.Urbakh$^3$}

\address{$^1$ Laboratoire de Physique Th\'eorique de la Mati\`ere Condens\'ee, Universit\'e Paris 6, Tour 24, 4 place Jussieu, 75252 Paris Cedex 05, France}

\address{$^2$ Max-Planck-Institut f\"ur Metallforschung, Heisenbergstr. 3,
D-70569 Stuttgart, Germany, and Institut f\"ur Theoretische und
Angewandte Physik, Universit\"at Stuttgart, Pfaffenwaldring 57,
D-70569 Stuttgart, Germany}

\address{$^3$ School of Chemistry, Tel Aviv University, 69978 Tel Aviv, Israel}

\ead{oshanin@lptl.jussieu.fr; klafter@post.tau.ac.il; urbakh@post.tau.ac.il}

\begin{abstract}
Dynamics of a classical particle in a one-dimensional, randomly driven
potential is analysed both analytically and numerically.
The potential considered here
is
composed of two identical spatially-periodic saw-tooth-like components,
one of which
 is externally driven
by a random force.
We show that under certain conditions
the particle may travel against the averaged external force
performing a saltatory unidirectional drift
with a constant velocity. Such
a behavior persists also in situations when the external force averages out to zero.
We demonstrate that the physics
behind this phenomenon stems from a particular behavior of fluctuations in random force: upon reaching a certain level, random fluctuations
exercise a locking function creating points of irreversibility
which the particle can not overpass. Repeated (randomly) in each
cycle, this results in a saltatory unidirectional drift.
This mechanism resembles the work of an escapement-type device in
watches. Considering the
overdamped limit, we propose simple analytical estimates for the
particle's terminal velocity.
\end{abstract}



\maketitle

\section{Introduction}

A question of how to make use of unbiased random or periodic
fluctuations rectifying them into directional motion has been a
long-standing challenge for mechanists, physicists and biologists.
For fluctuations acting on the microscopic level, originating, e.g.,
from Brownian noise, this problem has been under debate since the
early works of Maxwell and Smoluchowski \cite{maxwell,smoluchowski},
who formulated several important concepts (see also
Ref.\cite{Feynman/Leighton/Sands:1963}). In particular, it has been
realized already by Curie more than hundred years ago
\cite{Curie:1894}, that although the violation of the $x \rightarrow
-x$ symmetry is not sufficient to cause a net directional transport
of a particle subject to a spatially asymmetric, but on large scale
homogeneous, potential, the additional breaking of time reversal $t
\rightarrow -t$ symmetry, (e.g., due to dissipation), may give rise
to a macroscopic net velocity. Thus, in principle, on the
microscopic scale directed drift motion may emerge in the absence of
any external net force. These systems, referred nowadays to as
thermal ratchets \cite{Feynman/Leighton/Sands:1963}, have been
scrutinized both theoretically
(see, e.g. Refs.
[5-16]) and experimentally (see, e.g. Refs.[17-20]).
Considerable progress in this field, useful concepts and important
results have been summarized in several extensive reviews [21-24].

As a matter of fact, on the macroscopic scale of our everyday life
this very problem of obtaining useful work from random or periodic
perturbations has been and continues to be successfully accomplished
by engineering some clever technical devices; water- or windmills
and watches being just a few stray examples. In particular, to make
the watches work, the watchmakers had to invent a device capable to
convert the raw power of the driving force into regular and uniform
impulses, which was realized by creating various so-called
\textit{escapement} devices (see, e.g., an exposition in the web-site
\cite{watch}). In practice, this is most often a sort of a shaft or
an arm carrying two tongues - pallets, which alternately engage with
the teeth of a crown-wheel. The pallets follow the oscillating
motion of the controller - the balance-wheel or a pendulum, and in
each cycle of the controller the crown-wheel turns freely only when
both pallets are out of contact with it. Upon contacts, the pallets
provide impulses to the crown-wheel, (which is necessary to keep the
controller from drifting to a halt), and moreover, perform a locking
function stopping the train of wheels until the swing of the
controller brings round the next period of release.

In our recent paper \cite{wir} we have demonstrated that the simple
model, originally proposed in Ref.\cite{Porto/Urbakh/Klafter:2000},
works precisely like such an escapement device. This model consists
of a classical particle in a one-dimensional two-wave potential
composed of two periodic in space, identical time-independent
components, one of them being translated with respect to the other
by some external force.  It was found in
Ref.\cite{Porto/Urbakh/Klafter:2000} that for a constant or periodic
in time external driving force, the particle can perform directed
motion with a constant velocity. In our previous work \cite{wir} we
have realized that such unidirectional drift motion persists also in
situations when the driving force is random and averages out to
zero. We have shown that when the \textit{direction} of the driving
force fluctuates randomly in time,
 upon reaching a certain level,  random
fluctuations exercise a locking function creating points of
irreversibility which the particle can not overpass.
 Repeated in each cycle,
this process ultimately results in a saltatory drift motion with
random pausing times.

Here we consider the case of \textit{randomly directed} force in
more detail; in particular, we generalize our previous analysis for
systems in which the averaged force may have negative or positive
non-zero values. Focussing on the overdamped limit, we exploit the
mapping proposed in Ref.\cite{wir}, in which the original dynamical
system was viewed as a biased Brownian motion on a hierarchy of
disconnected intervals, and derive simple analytical estimates for
the particle's terminal velocity. These results generalize our
previous findings and demonstrate that under certain conditions the
particle may even perform unidirectional saltatory drift motion
against the direction of the averaged external force.

\section{The Model}

Consider a periodic, piecewise continuous function $\Pi(x)$ (see,
Fig.1). Without lack of generality, we assume that the periodicity
$b$ of this function is equal to unity, $b = 1$, so that $\Pi(x+1) =
\Pi(x)$ for any $x$, and an amplitude $\Pi_0 = \max \Pi(x) = -\min
\Pi(x)$. Within each period the function $\Pi(x)$ has a saw-tooth
shape:
\begin{equation}\label{eq:ratchet}
\Pi(x) \equiv \Pi_0 \left\{
\begin{array}{ll}
\displaystyle -1 + 2 x/\xi &
\mbox{if $\displaystyle x \le \xi$}\\[5mm]
\displaystyle 1 - 2 (x- \xi)/(1 - \xi) &
\mbox{if $\displaystyle x > \xi$}
\end{array}\right. \quad
\end{equation}
 The parameter
$\xi \in (0, 1 )$ determines the asymmetry of the saw-tooth, with
$\xi = 1/2$ corresponding to the symmetric case. We focus here on
the case $\xi < 1/2$.

\begin{figure}\centering
\includegraphics[width=10cm]{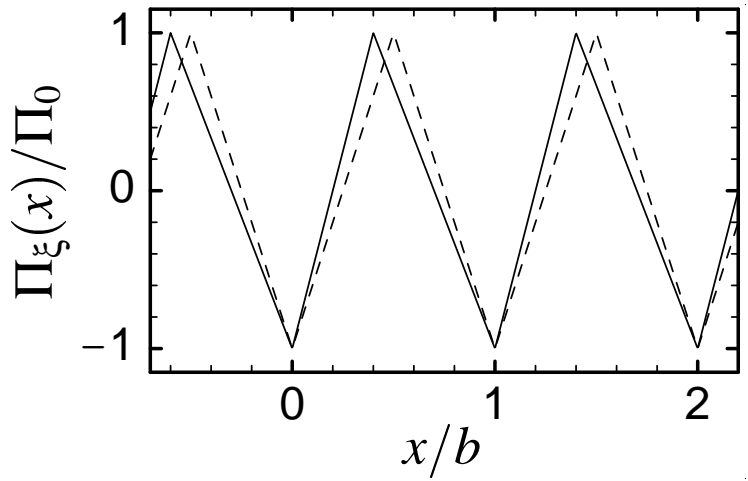}
\caption{Plot of the function given by Eq.~(\ref{eq:ratchet}) with
asymmetry parameters $\xi = 2/5$ (solid line) and $\xi' = 1/2$
(dashed line, symmetric case).}
\end{figure}

Following the model of Ref.\cite{Porto/Urbakh/Klafter:2000}, we construct now
the potential $V(x,\gamma_t)$ as the sum
 $V(x,\gamma_t) \equiv \Pi(x) + \Pi(x-\gamma_t)$,
where $\gamma_t$ defines the external translation. Note that since
both $\Pi(x)$ and $\Pi(x-\gamma_t)$ are periodic, the potential
$V(x,\gamma_t)$ is periodic in both arguments, so that
$V(x,\gamma_t+1) = V(x,\gamma_t)$ for any $\gamma_t$.

Consider now a classical particle of mass $m$ moving in the potential
$V(x,\gamma_t)$.
The deterministic equation of motion in this potential landscape reads:
\begin{equation}\label{eq:motion}
\fl
m \ddot{x}_t + \eta \dot{x}_t + \frac{\partial V(x_t,\gamma_t)}{\partial x_t} = 0,
\end{equation}
where $x_t$ denotes the particle's trajectory. Note that energy is
steadily pumped into the system through the external translation
$\gamma_t$. On the other hand, the energy is steadily
dissipated when the damping $\eta > 0$,
which prevents the particle from detaching from the system.

Eventually, we suppose that the translation $\gamma_t$ originates
from an external random force $f_t$. More specifically, we stipulate
that the translation $\gamma_t$ obeys the Langevin equation,
\begin{equation}
\fl
\dot{\gamma_t} = f_t,
\end{equation}
$f_t$ being a random Gaussian, delta-correlated process with
moments:
\begin{equation}
\label{force}
\fl
\overline{f_t} = f_0, \;\;\; \overline{f_t f_{t'}} = f_0^2 + 2 D \delta(t - t'),
\end{equation}
where the overbar denotes averaging over thermal histories.

\section{Particle Dynamics in the Overdamped Limit}

We will restrict ourselves to the limit of an overdamped motion,
$\eta/[(2 \pi) \; \sqrt{m \Pi_0}] \gg 1$ and suppose that initially, at $t = 0$,
$x_0 = 0$ and $\gamma_0 = 0$, so that the particle is located
at a potential minimum.
Now, in the overdamped limit, as shown in Ref.\cite{Porto/Urbakh/Klafter:2000},
 the particle dynamics
is governed entirely by the evolution of the
minima of the total potential $V(x,\gamma_t)$. Consequently, in
order to obtain  the trajectory $x_t$ and estimate the statistical
velocity $V = d\overline{x_t}/dt$,
 it suffices to study the time evolution of
positions of these minima. This process has been discussed in
Ref.\cite{Porto/Urbakh/Klafter:2000} and we address the reader to
this work for more details. Here we outline the main conclusions
of Ref.\cite{Porto/Urbakh/Klafter:2000}: The potential
$V(x,\gamma_t)$ possesses a set of minima and the position of each
minimum changes in time as the translation $\gamma_t$ evolves. The
particle, located at $t = 0$ in the first minimum simply follows
the motion of this minimum up to a certain time when it reaches
the point $x = \tilde{x}$. The $\tilde{x}$-points
 are points of instability ${\cal I}$ in the $(x,\gamma_t)$ plane, (emerging
due to the asymmetry of functions $\Pi(x)$, $\zeta < 1/2$), where the
corresponding local minimum of the potential $V(x,\gamma_t)$
ceases to exist. As soon as the minimum at which the particle
resides disappears, the particle performs an irreversible jump,
instantaneous on the time scale of the translation, to one of the
two still existing neighboring minima. In the overdamped limit,
the direction of the jump is prescribed by the slope of
$V(x,\gamma_t)$ at this point. The particle moves with the second
minimum until it ceases to exist and then jump to the next minimum
and so on.

\section{Evolution of the effective translation $\Gamma_t$.}

The topology of the set ${\cal I}$ formed by
the points of instability in the $(x,\gamma_t)$ plane has been amply analysed in
Ref.\cite{Porto/Urbakh/Klafter:2000} (see, Figs.1 and 2 and the discussion below)
on example of a particle subject to the potential in Eq.(1), one component of which experiences a \textit{constant} translation. It has been realized 
that the set ${\cal I}$ comprises special points $-K+\xi$ and $-K + 3/2$, $K = 1,2, \ldots$, 
at which, depending on the direction at which each of these points is approached by the particle, 
particle trajectoriy experiences some irregularities.  

To illustrate these irregularities in particle dynamics, 
it is expedient to introduce
\textit{an effective} translation $\Gamma_t$
as a function obeying formally the following Langevin-type equation
\begin{equation}
\label{L}
\fl
\dot{\Gamma}_t = \hat{\cal L}\left(\Gamma_t\right) + f_t,
\end{equation}
where the moments of a random Gaussian force $f_t$ have been defined
in Eq.(\ref{force}), while $\hat{\cal L}\left(\Gamma_t\right)$ is an
operator, which equals zero everywhere except of the set of special
points $\Gamma_t = - K + \xi$ and $\Gamma_t = - K + 3/2$, $K = 1, 2,
\ldots$.

\begin{figure}\centering
\includegraphics[width=12cm]{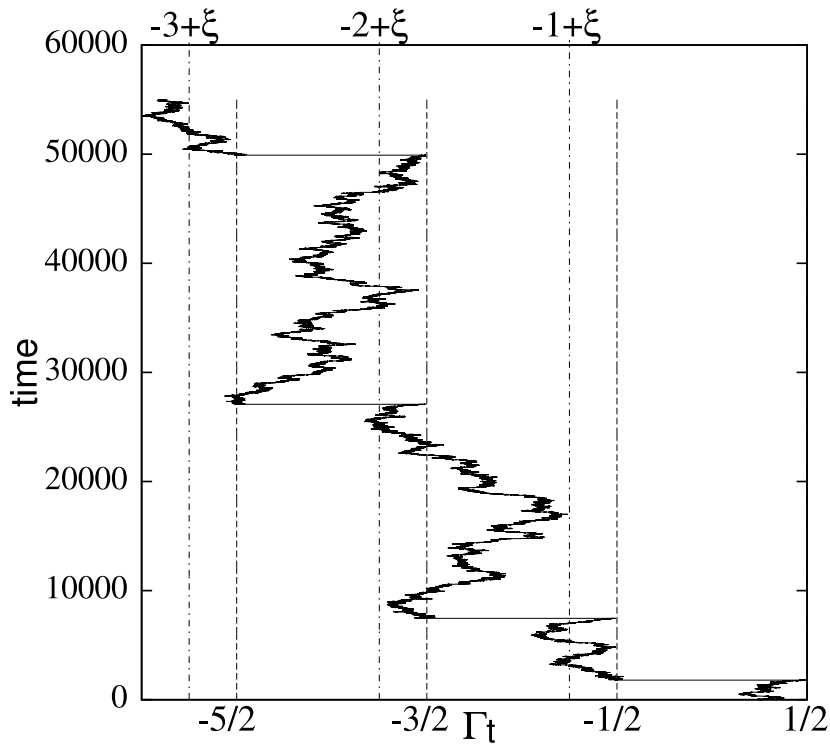}
\caption{Typical realization of the process $\Gamma_t$. The dashed lines
denote the "reflection points" $-K + 3/2$ of the trajectory, while
the dash-dotted lines - the points $-K + \xi$.}
\end{figure}

At these special points, the action of the operator $\hat{\cal
L}\left(\Gamma_t\right)$ is as follows: When  $\Gamma_t$ reaches the
value $\Gamma_t = - K + 3/2$, (which we will call the "reflection
point"), the operator $\hat{\cal L}\left(\Gamma_t\right)$ changes
$\Gamma_t \to \Gamma_t - 1$, i.e. shifts instantaneously its value
from $\Gamma_t = - K + 3/2$ to  $\Gamma_t = - K + 1/2$. On the other
hand, when  $\Gamma_t$ hits the value $ - K + \xi$, the operator
$\hat{\cal L}\left(\Gamma_t\right)$ shifts the position of the
reflection point at  $- K + 3/2$ to $- K + 1/2$.

In Fig.2 we depict a typical realization of the process $\Gamma_t$,
defined by Eqs.(\ref{L}) and (\ref{force}) in case when the force
averages out to zero, $f_0 = 0$. In numerical simulations of
$\Gamma_t$, we discretised the process defined by Eq.(\ref{L})
taking the lattice spacing $\delta \Gamma$ equal to $2 \; \times \;
10^{-2}$ and the characteristic jump time $\tau$ has been set equal
to unity. Hence, the diffusion coefficient $D = (\delta \Gamma)^2/2
\tau =  2 \; \times \; 10^{-4}$.

Note that the operator $\hat{\cal L}\left(\Gamma_t\right)$
encompasses all essential physics associated with the ratchet
effect; due to the presence of the operator $\hat{\cal
L}\left(\Gamma_t\right)$ the process $\Gamma_t$ experiences an
effective drift in the negative direction (for $\xi < 1/2$); in
absence of $\hat{\cal L}\left(\Gamma_t\right)$ one should, of
course, obtain purely diffusive behavior such that
$\overline{\Gamma_t} \equiv 0$ and $\overline{\Gamma^2_t} \equiv 2 D
t$.

Now, the process $\Gamma_t$ defined by Eq.(\ref{L}) can be viewed
from a different perspective:
 namely, $\Gamma_t$ can be regarded formally
as a biased (for $f_0 \neq 0$) Brownian motion evolving in time $t$
on a hierarchical lattice composed of a semi-infinite set of
independent intervals of length $3/2 - \xi$ (see Fig.3). The
Brownian motion starts at $t = 0$ at the origin of the first
interval ($K = 1$) and evolves freely until it either hits the
right-hand-side boundary $\Gamma_t = 1/2$ (which is called, in what
follows, the "reflection point"), in which case it gets transferred
instantaneously at position $\Gamma_t = - 1/2$ and continues its
motion on the interval $K = 1$, or reaches the left-hand-side
boundary $\Gamma_t = - 1 + \xi$ - the "exit point" and gets
irreversibly transferred to the second ($K = 2$) interval. In the
second, and etc, interval the process $\Gamma_t$  evolves according
the same rules.

\begin{figure}\centering
\includegraphics[width=10cm, angle=-90]{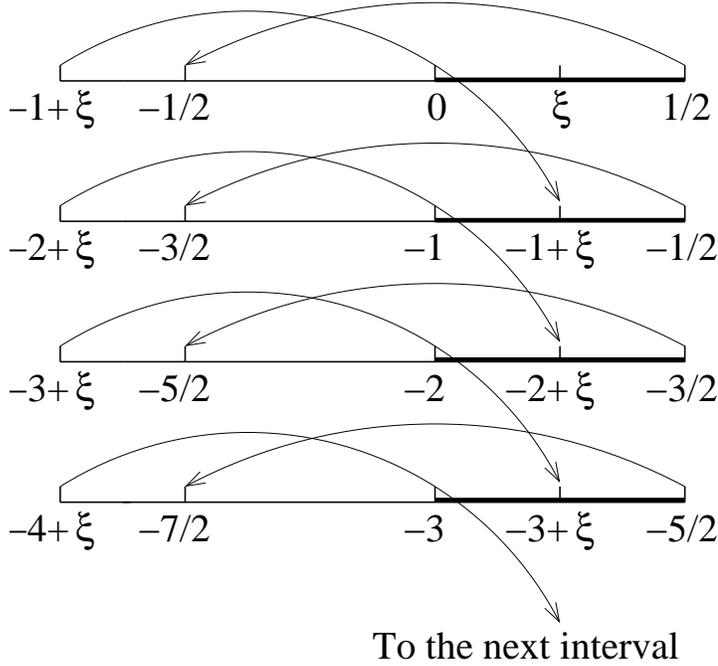}
\caption{Time evolution of $\Gamma_t$ viewed as a diffusive process on a
hierarchy of disconnected intervals.}
\end{figure}

The trajectory of
the particle $x_t$, a particular realization of which we depict in Fig.4,
follows essentially the process $\Gamma_t$,
i.e. $x_t = \Gamma_t$,
except for the following two
situations:\\
(i) when $\Gamma_t$ appears within the intervals $[- K + 1, - K +
3/2]$, $K = 1, 2, \ldots$, (thick lines in Fig.3), the particle
position stops on the left boundary of these intervals, i.e., on a
given interval (a given $K$) $  x_t \equiv - K + 1$  when $\Gamma_t$
is within the interval $[- K + 1, - K + 3/2]$. Therefore, the
particle seizes to move for some time, this "pausing" time being
equal to the time which $\Gamma_t$ spends within this interval. The
overall trajectory $x_t$ is thus highly saltatory.

\begin{figure}\centering
\includegraphics[width=10cm, angle=-90]{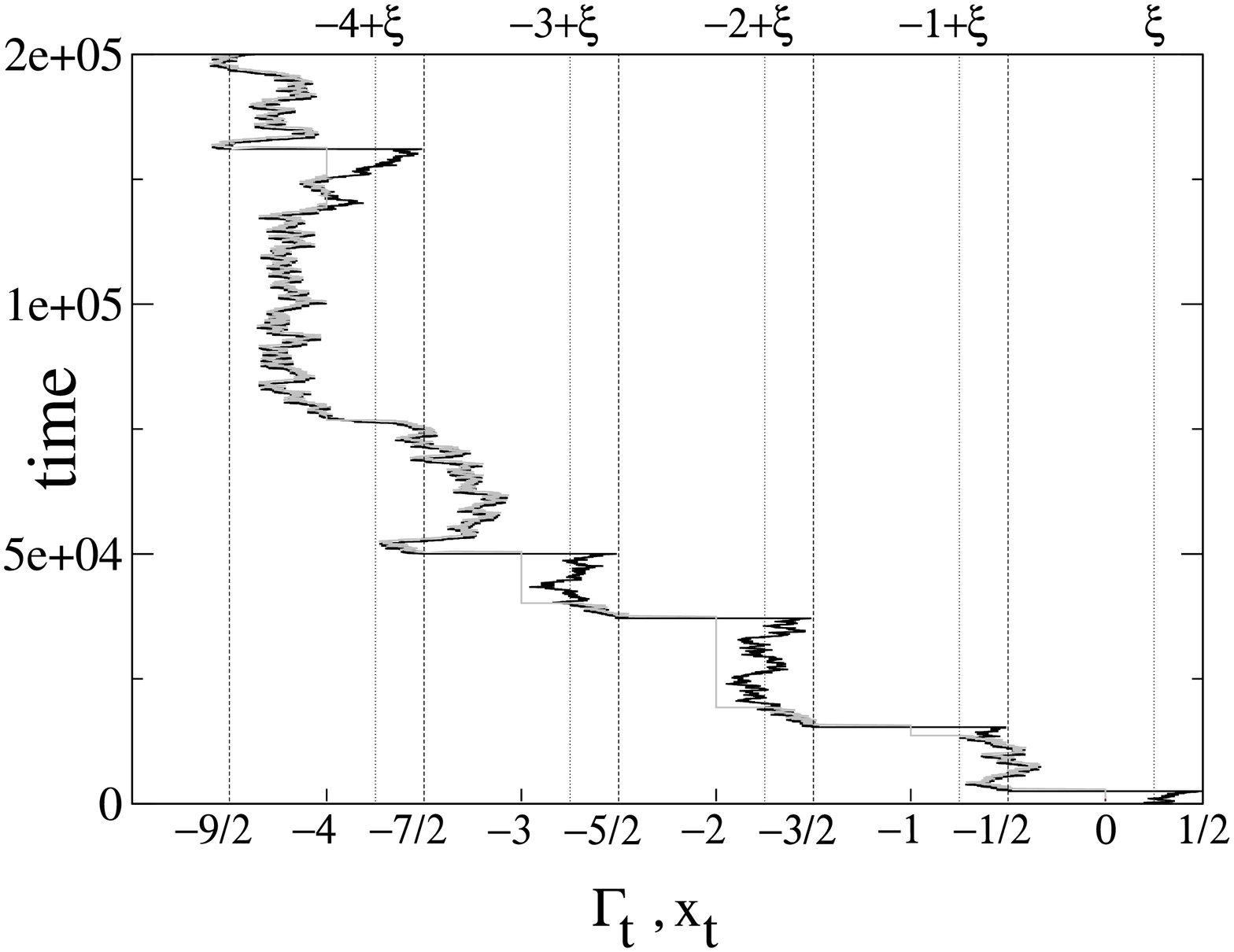}
\caption{ Typical realization of the effective translation
$\Gamma_t$ (black line) and the corresponding particle trajectory
$x_t$ (gray line). The dashed lines denote the "reflection points"
$x=-K + 3/2$ of the trajectory $\Gamma_t$, while the dash-dotted
lines - the special points $-K + \xi$.  Vertical red lines show the
special points at which the particle seizes to move.}
\end{figure}

(ii) when $\Gamma_t$ hits the left-hand-side boundary of the $K$-th
interval and gets instantaneously transferred to the next interval,
the particle makes a jump from $x_t = - K +\xi$ to $x_t = - K$. It
means that although the trajectory $\Gamma_t$ is always
continuous, the particle trajectory $x_t$ changes discontinuously
at these points. It is clear, however, that such jumps of length
$\xi$ do not contribute to the average particle velocity $
\overline{\dot{x}_t}$. Consequently,  the velocity of the effective
external translation ${\cal V} = \overline{\dot{\Gamma}_t}$ and
average particle velocity $\overline{\dot{x}_t}$ coincide, i.e.
${\cal V} = \overline{\dot{x}_t} = \overline{\dot{\Gamma}_t}$.

\section{Zero external bias $f_0$.}

We turn next to the computation of the velocity   ${\cal V}$ focussing
first on the case of zero external bias, $f_0 = 0$. Evidently, even in
this simplest case the map depicted in Fig.3 is too complex
to be solved exactly. However, some simple analytical arguments can be
proposed to obtain   very accurate estimates for the functional form of
 ${\cal V}$. To do this, we first note that it is only when passing
from the $K$-th interval to the next, $(K+1)$-th interval, (which is
displaced on a unit distance on the $\Gamma_t$-axis with respect to
the previous one), the process $\Gamma_t$ gains an uncompensated
(negative) displacement equal to the overlap distance of two
consecutive intervals, which equals, namely, $1/2 - \xi$.
Consequently,  ${\cal V}$ can be estimated as  ${\cal V} = - (1/2 -
\xi)/T$, where $T$ is the mean time which the process $\Gamma_t$
"spends" within a given interval. To estimate the value of $T$, let
us consider more precisely dynamics of the process $\Gamma_t$ on a
given interval.

\begin{figure}\centering
\includegraphics[width=12cm]{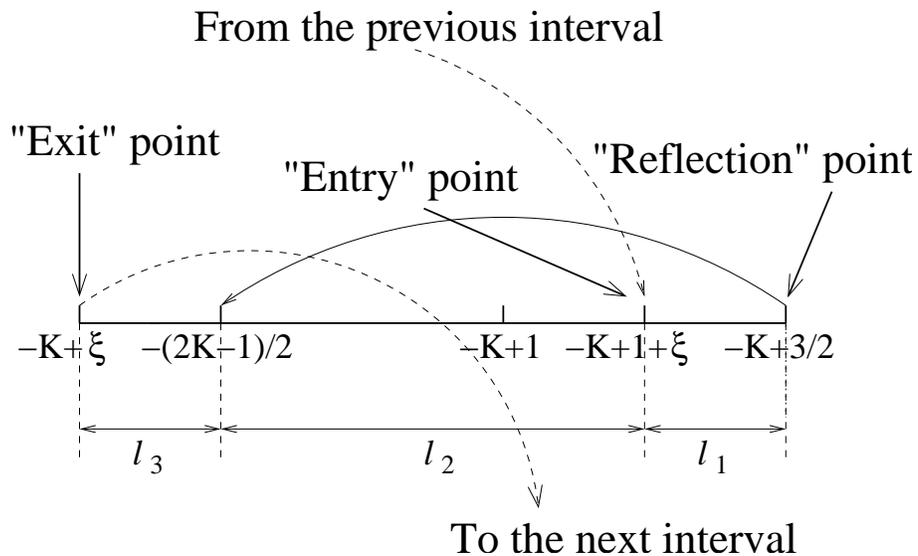}
\caption{
Time evolution of the effective external translation $\Gamma_t$
on the $K$-th interval.}
\end{figure}

The time $T$ is, as a matter of fact, the mean time needed for the
process $\Gamma_t$ to reach diffusively for the first time the
left-hand-side boundary of the interval - the "exit" point $- K +
\xi$, starting from the "entry" point $-K + 1 + \xi$ (see Fig.5). It
is convenient now to divide the $K$-th interval into three
sub-intervals: $l_1 = l_3 = 1/2 - \xi$ and $l_2 = 1/2 + \xi$, Fig.5.
Evidently, since $l_1 = l_3 \leq l_2$, ($\xi > 0$), for typical
realizations of external random force $f_t$, the following scenario
holds: the process $\Gamma_t$ evolves randomly around the "entry"
point and first hits the right-hand-side boundary of the $K$-th
interval passing thus through the sub-interval $l_1$; the mean time
required for the first passage of this sub-interval is denoted as
$T_1$. For conventional diffusion with diffusion coefficient $D$
this time is given by $T_1 = (1/2 - \xi)^2/2 D$ \cite{1}. Further
on, after hitting the right-hand-side boundary,  $\Gamma_t$ gets
reflected to the point $\Gamma_t = - (2 K - 1)/2$ and evolves
randomly around this point until it hits the left-hand-side boundary
of the interval - the "exit" point. For symmetric diffusion, the
time $T_3$ needed for the first passage through the sub-interval
$l_3$ is equal to $T_1$. Consequently, $T = T_1 + T_3 =   (1/2 -
\xi)^2/D$ and the particle velocity follows
\begin{equation}
\label{velocity}
\fl
{\cal V} = - \frac{2 D}{1 - 2 \xi}
\end{equation}
In Fig.6 we compare our theoretical prediction in
Eq.(\ref{velocity}) and the results of the Monte Carlo simulations,
which comparison shows that simple theoretical arguments presented
above capture the essential physics underlying the dynamics of the
random map depicted in Fig.3.

\begin{figure}\centering
\includegraphics[width=12cm]{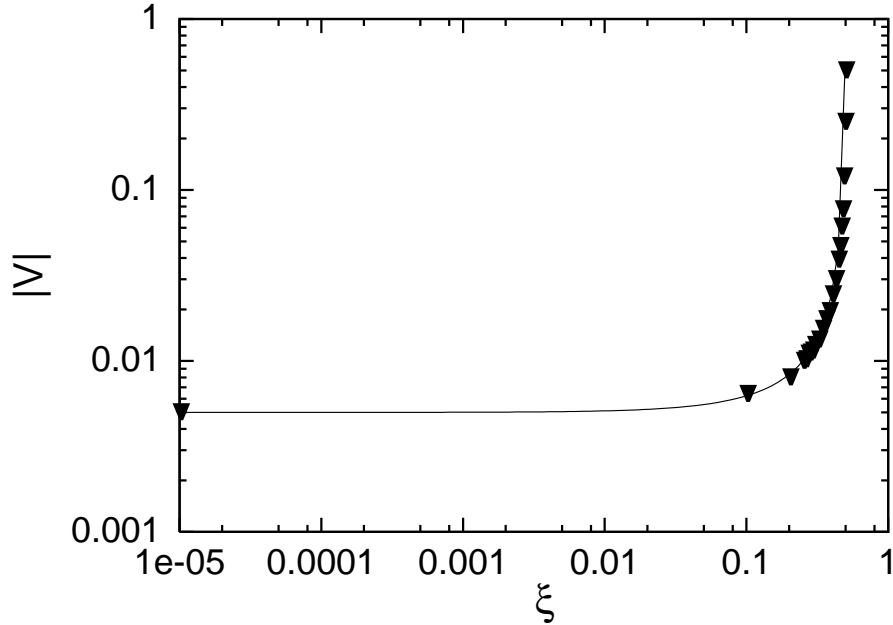}
\caption{The absolute value of the statistical velocity $V$ versus
$\xi$. The solid line gives the analytical prediction in
Eq.(\ref{velocity}) and the triangles depict Monte Carlo data. Note
that we use, for notational convenience, the log-log scale.}
\end{figure}

Now, we proceed with several comments on the result in
Eq.(\ref{velocity}).\\
(i) First, we note that  ${\cal V}$ determined by
Eq.(\ref{velocity}) increases indefinitely when $\xi \to 1/2$,
although, however, ${\cal V} \equiv 0$ for $\xi \equiv 1/2$, which
is a seemingly artificial discontinuity. As a matter of fact, such a
behavior stems from our definition of the random process $\Gamma_t$;
namely, by writing our
 Eq.(\ref{L}) we tacitly assume that both the
effective step $\delta \Gamma$ of the process $\Gamma_t$ and the
characteristic jump time $\tau$ are infinitesimal variables, (while
the ratio $D = (\delta \Gamma)^2/2 \tau$ is supposed to be fixed and
finite). On the other hand, for any realistic physical system
$\delta \Gamma$ and $\tau$ may be very small, but nonetheless are
both finite. For finite $\delta \Gamma$ and $\tau$, the process
$\Gamma_t$ can be viewed as a hopping process on a lattice with
spacing $\delta \Gamma$ and with the characteristic time separating
successive jumps equal to $\tau$. The hopping probabilities are
symmetric; that is, the probabilities of jumping from the site
$\Gamma$ to either of its neighboring sites are equal to each other.
For this model, we find for $T_1$ and $T_3$ the following forms
(see, e.g., Ref.\cite{1}):
\begin{equation}
\label{T}
\fl
T_1 = T_3 = \tau \Big(1 + L\Big) L,
\end{equation}
where $L = (1/2 - \xi)/\delta \Gamma$ is the number of
 elementary steps $\delta \Gamma$
in the interval $(1/2 - \xi)$.
Equation (\ref{T})
yields, in place of Eq.(\ref{velocity}),
\begin{equation}
\label{velocityfin} \fl {\cal V} = - \frac{2 D}{2 \delta \Gamma + 1
- 2 \xi}
\end{equation}
Note now that for finite $\delta \Gamma$ and $\tau$ the velocity
 ${\cal V}$ in Eq.(\ref{velocityfin})
tends to a finite value when $\xi \to 1/2$ (but still we have ${\cal
V} \equiv 0$ for $\xi = 1/2$). We also remark that relaxing the
assumption of the overdamped motion and/or adding a noise term to
Eq.(\ref{eq:motion}), we will obtain a certain rounding of the
behavior of ${\cal V}$ in the vicinity of $\xi  = 1/2$ \cite{2}. As
a matter of fact, our numerical simulations show that ${\cal V}$
becomes a smooth bell-shaped function with ${\cal V} = 0$ for $\xi =
1/2$ and a maximal value close to $\xi = 1/2$.

(ii) Second, we notice that the velocity defined by
Eqs.(\ref{velocity}) or (\ref{velocityfin}) is a monotonously
increasing function of the parameter $\xi$, which defines the degree
of the asymmetry of the external potential; that is, the velocity by
absolute value is maximal for low asymmetry when $\xi \to 1/2$ and
minimal, ${\cal V} = 2 D$, for the strongest asymmetry with $\xi =
0$. We would like to comment now that such a behavior relies
strongly on the definition of the random external force $f_t$ as a
Gaussian, delta-correlated noise. As $f_t$ is the property
influenced by external processes, it might be characterized, for
instance, by essential correlations or be a non-local in space or in
time (discontinuous) stochastic process. In both cases, the behavior
of the particle's velocity in the presence of such an external force
would be different of that predicted by Eqs.(\ref{velocity}) and
(\ref{velocityfin}).

Suppose, for example, that $f_{t}$ represents the so-called
delta-correlated L\'evy  process with parameter $\mu$ (see, e.g.,
Ref.\cite{1} for more discussion). We note here parenthetically that
the case $\mu = 2$ corresponds to the usual Gaussian case, which
yields conventional diffusive motion, while
 the case $\mu = 1$ describes the so-called Cauchy process.

Now, what basically changes when we assume that $f_t$ represents the
L\'evy process is the functional form of the first passage times
$T_1$ and $T_2$. In this case the time required for the first
passage of an interval of length $1/2 - \xi$ reads $T_1 = T_3 \sim
(1/2 - \xi)^\mu$ \cite{1} and hence, the particle velocity follows
\begin{equation}
\label{velocitylevy}
\fl
{\cal V} \sim - (1 - 2 \xi)^{1 - \mu}
\end{equation}
Consequently, we may expect that the particle's velocity will be a
monotonously increasing function of the parameter $\xi$ only for the
L\'evy processes with $\mu > 1$, i.e. for persistent processes
(which have a finite probability for return to the origin). In the
borderline case of $\mu = 1$ (the Cauchy process) velocity will be
independent of the asymmetry parameter. On the other hand, for
transient processes with $\mu < 1$, (i.e. having zero probability of
return to the origin), which are not space-filling (fractal) and
occupy the space in clustered or localized patches, one may expect
that the particle's velocity would be a \textit{decreasing} function
of the asymmetry parameter $\xi$.

\section{Non-zero external bias $f_0$.}

 In this section we generalize our previous results for the generic
 case of non-zero averaged external force. We will consider both the
 situations when $f_0$ is negative or positive.

\subsection{ Positive bias}

Consider first the case when the bias is oriented in the positive
direction, i.e. $f_0 \geq 0$. The salient feature of this situation
is that here the asymmetry of the embedding potential prevents the
particle of moving in the positive direction, i.e. in the direction
of the applied field $f_0$. Hence, in this case if a drift motion of
the particle takes place, - it would take place \textit{against} the
applied field! We set out to show in what follows that this is
actually the case: the particle does travel with a constant velocity
${\cal V}$ in the negative direction and such an extraordinary
behavior is essentially due to fluctuations in the process
$\Gamma_t$ - some (exponentially small) number of its trajectories
traveling against the external field.

Now, we suppose that the hopping process $\Gamma_t$ takes place on a
discrete lattice with a spacing $\delta \Gamma$. Each interval in
Fig.5 is thus separated into a set of $L$ elementary steps, $L =
(3/2 - \xi)/\delta \Gamma$. The hopping probabilities for jumps in
the positive ($p$) and in the negative ($q$) directions are no
longer equal and obey:
\begin{equation}
\fl
p/q = \exp\Big(\beta f_0 \delta \gamma\Big), \;\;\; p + q = 1,
\end{equation}
where $\beta$ denotes the reciprocal temperature.

One readily notices now that in the situation under consideration
 the symmetry $T_1 = T_3$, which existed in the case $f_0 = 0$,  is broken: when
passing through the sub-interval $l_1$ the process $\Gamma_t$
follows the field, while the first passage through the sub-interval
$l_3$ takes place in the situation when the field $f_0$ is directed
effectively against the passage direction. Supposing that $\delta
\Gamma$ and $\tau$ are both finite, we have then that here \cite{1}:
\begin{eqnarray}
\label{T1} \fl T_1 = \frac{\tau \Big(1 + \exp\Big(- \beta f_0 \delta
\Gamma\Big)\Big)}{\Big(1 -
 \exp\Big(- \beta f_0 \delta \Gamma\Big)\Big)} \Big[
\frac{1/2 - \xi}{\delta \Gamma} - \nonumber\\
- \frac{\exp\Big(- \beta f_0 \delta \Gamma\Big)}{\Big(1 -
 \exp\Big(- \beta f_0 \delta \Gamma\Big)\Big)} \Big(1 - \exp\Big(- \beta f_0
(1/2 - \xi)\Big)\Big)
\Big]
\end{eqnarray}
and
\begin{eqnarray}
\label{T3} \fl T_3  = \frac{\tau \Big(1 + \exp\Big(\beta f_0 \delta
\Gamma\Big)\Big)}{\Big( \exp\Big(\beta f_0 \delta \Gamma\Big) -
1\Big)} \Big[
 \frac{\exp\Big(\beta f_0 \delta
\Gamma\Big)}{\Big(\exp\Big(\beta f_0 \delta \Gamma\Big) - 1\Big)}
\times \nonumber\\ \times \Big(\exp\Big(\beta f_0 (1/2 - \xi)\Big) -
1\Big) - \frac{1/2 - \xi}{\delta \Gamma} \Big],
\end{eqnarray}
respectively.

Note that $T_1$ grows linearly with the sub-interval length $1/2 -
\xi$, while $T_3$ shows much stronger, \textit{exponential}
interval-length dependence, and hence, controls the overall time
spent within a given interval.

Consequently, the particle's velocity in this case attains the form
\begin{equation}
\label{velocityf} \fl {\cal V}  = - \frac{(1/2 - \xi)
\sinh^2\Big(\beta f_0 \delta \Gamma/2\Big)}{\tau \cosh\Big(\beta f_0
\delta \Gamma/2\Big) \Big[ \cosh\Big(\beta f_0 (1 - 2 \xi + \delta
\Gamma)/2\Big) -  \cosh\Big(\beta f_0 \delta \Gamma/2\Big)\Big]}
\end{equation}
We note again that the remarkable feature of this results is that it
predicts a drift motion of the particle \textit{against} the applied
external field $f_0$.

In the diffusion limit, i.e. when $\delta \Gamma$ and $\tau$ are set
equal to zero but the ratio $(\delta \Gamma)^2/\tau$ is kept fixed,
we find from Eq.(\ref{velocityf}) the following result:
\begin{equation}
\label{velocityfdif}
\fl
{\cal V} = - \frac{(1 - 2 \xi) D \beta^2 f_0^2 }{8
\sinh^2\Big(\beta f_0 (1 - 2 \xi)/4\Big)},
\end{equation}
which reduces to the result in Eq.(\ref{velocity}) when $f_0 \to 0$.

\subsection{Negative bias}

We turn next to the case of negative bias, $f_0 < 0$. At the first
glance, this case should not give us any surprise - one anticipates
that the particle here will always be passively carried by the
field. It appears, however, that such a behavior may take place only
for sufficiently low asymmetries of the saw-tooth potential. For
high asymmetries $\xi$, it may appear that it will be more efficient
for the particle to travel through the shorter interval ($l_1$,
Fig.5) against the field, than to be carried passively through the
longer interval $l_2$. It means that exponentially small number of
trajectories travelling against the field will still matter, as in
the case of positive bias. Consequently, this will give rise to a
cusp in the overall dependence ${\cal V}(\xi)$.

We turn now to the analysis of the characteristic passage times
involved in this case. We start first with the passage through the
sub-interval $l_3$, in which a particle appearing at some time at
the point $-(2 K - 1)/2$ should travel to the "exit" point $-K +
\xi$. The passage through this sub-interval is favored by the field
$f_0$ and the first passage time is given by
\begin{eqnarray}
\label{T33} \fl T_3 = \frac{\tau \Big(1 + \exp\Big(\beta f_0 \delta
\Gamma\Big)\Big)}{\Big(1 -
 \exp\Big(\beta f_0 \delta \Gamma\Big)\Big)} \Big[
\frac{1/2 - \xi}{\delta \Gamma} - \nonumber\\ -\frac{\exp\Big(\beta
f_0 \delta \Gamma\Big)}{\Big(1 -
 \exp\Big(\beta f_0 \delta \Gamma\Big)\Big)} \Big(1 - \exp\Big(\beta f_0
(1/2 - \xi)\Big)\Big)
\Big]
\end{eqnarray}
Now, the question what is the mean time typically needed to reach this point
$-(2 K - 1)/2$
starting from the "entry"  point $-K + 1 + \xi$ is a bit more delicate: it can be
reach either by the field-enhanced diffusion through the larger sub-interval $l_2$ or
by diffusion against the field $f_0$ through the smaller sub-interval $l_1$
to the right-hand-side boundary of the
interval (the "reflection" point) followed by an instantaneous transfer
to the point $-(2 K - 1)/2$ (see, Fig.3).
Two respective mean first passage times are given here by
\begin{eqnarray}
\fl T_2 = \frac{\tau \Big(1 + \exp\Big(\beta f_0 \delta
\Gamma\Big)\Big)}{\Big(1 -
 \exp\Big(\beta f_0 \delta \Gamma\Big)\Big)} \Big[
\frac{1/2 + \xi}{\delta \Gamma} - \nonumber\\ - \frac{\exp\Big(\beta
f_0 \delta \Gamma\Big)}{\Big(1 -
 \exp\Big(\beta f_0 \delta \Gamma\Big)\Big)} \Big(1 - \exp\Big(\beta f_0
(1/2 + \xi)\Big)\Big)
\Big]
\end{eqnarray}
and
\begin{equation}
\fl T_1 = \frac{\tau \Big(1 + \exp\Big(\beta f_0 \delta
\Gamma\Big)\Big)}{\Big(1 -
 \exp\Big(\beta f_0 \delta \Gamma\Big)\Big)} \Big[
 \frac{\Big(\exp\Big(\beta f_0
(1/2 - \xi)\Big)\Big) - 1}{\Big(1 -
 \exp\Big(-\beta f_0 \delta \Gamma\Big)\Big)} - \frac{1/2 - \xi}{\delta \Gamma}
\Big]
\end{equation}
The particle velocity ${\cal V}$ thus can be estimated as:
\begin{equation}
\fl
{\cal V} = - \frac{(1/2 - \xi)}{T_3 + min(T_1,T_2)}
\end{equation}
If $T_1 < T_2$, we would evidently recover our previous result in
Eq.(\ref{velocityf}) up to the reversal of the sign, $f_0 \to -
f_0$. In the opposite case, $T_1 > T_2$, one finds
\begin{eqnarray}
\label{new} \fl {\cal V} = - \frac{(1/2 - \xi) \delta \Gamma}{\tau
\coth\Big(-\beta f_0 \delta \Gamma/2\Big)} \Big[1 - \frac{2 \delta
\Gamma \exp\Big(\beta f_0 \delta \Gamma\Big)}{1 - \exp\Big(\beta f_0
\delta \Gamma\Big)} \times \nonumber\\ \times \Big(1 -
\exp\Big(\beta f_0 \delta \Gamma/2\Big) \cosh\Big((\beta f_0 \delta
\Gamma \xi\Big)\Big) \Big]^{-1}
\end{eqnarray}

Now, on comparing $T_1$ and $T_2$ we find that these two mean first passage times are
equal to each other when the following equality holds:
\begin{equation}
\fl \xi = \xi_{crit} = \frac{\ln\Big(\cosh\Big(\beta f_0 \delta
\Gamma/2\Big)\Big)}{\Big(-\beta f_0 \delta \Gamma\Big)}
\end{equation}
Hence, for fixed $\beta f_0 \delta \Gamma$, $T_1 < T_2$ and the
particle velocity is described by Eq.(\ref{velocityf}) with $f_0$
changed for $-f_0$ when
\begin{equation}
\xi > \xi_{crit}
\end{equation}
Otherwise,  $T_2 < T_1$ and the particle velocity obeys
Eq.(\ref{new}). This means that the overall dependence of the
particle velocity on the asymmetry parameter will have a cusp at
$\xi = \xi_{crit}$. Note also that $\xi_{crit}$ is a monotonously
increasing function of the parameter $( - \beta f_0 \delta \Gamma)$,
which equals zero when $\beta f_0 \delta \Gamma = 0$ and rapidly
approaches
 $1/2$ when $( - \beta f_0 \delta \Gamma) \to \infty$. Hence, $\xi$ should be
sufficiently close to its maximal value $1/2$ in order that the behavior predicted by
Eq.(\ref{velocityf}) takes place in the case of a negative bias.

\section{Conclusion}

In conclusion,  we have studied dynamics of a classical particle in
a 1D potential, composed of two periodic, asymmetric saw-tooth
components, one of which is driven by an external random force.
Concentrating on the overdamped limit, we  presented analytical
estimates for the particle's velocity. We have shown that in such a
system the particle will perform a $unidirectional$ saltatory drift
regardless of the fact whether the average force is zero, or
directed against the particle motion. In case when the average force
is negative and its direction coincides with the direction of the
particle motion, we predicted a cusp-like dependence of the particle
velocity on the asymmetry parameter of the saw-tooth potential. We
have demonstrated that the physical mechanism underlying such a
behavior resembles the work of the so-called escapement device, used
by watchmakers to convert the raw power of the driving force into
uniform impulses. Here, indeed, upon reaching certain levels, random
forces lock the particle's motion creating points of
irreversibility, so that the particle gets uncompensated
displacements. Repeated (randomly) in each cycle, this process
ultimately results in a saltatory drift with random pausing times.
Our analytical results for systems in which the random force
averages out to zero are in a very good agreement with Monte Carlo
data.

\section{Acknowledgments}
The authors gratefully acknowledge helpful discussions with M.Porto.
GO thanks the AvH Foundation for the financial
support via the Bessel Research Award.

\end{document}